\documentclass[aps,pre,twocolumn]{revtex4-1}
\usepackage{amsmath,amssymb,graphicx}
\input epsf
\usepackage{amsbsy}
\usepackage{latexsym}
\usepackage{color}
\usepackage{graphicx}
\usepackage{psfrag}
\usepackage[normalem]{ulem}
\usepackage{float}

\usepackage{booktabs}

\newcommand{\be}{\begin{equation}}
\newcommand{\ee}{\end{equation}}

\begin{document}

\title{Quantitative and qualitative analysis of asynchronous neural activity}

\author{Ekkehard Ullner, Antonio Politi}
\affiliation{Institute for Pure and Applied Mathematics and Department of Physics (SUPA), Old Aberdeen, Aberdeen AB24 3UE, United Kingdom}
 
\author{Alessandro Torcini}
\affiliation{Laboratoire de Physique Th\'eorique et Mod\'elisation, Universit\'e de Cergy-Pontoise, CNRS, UMR 8089,95302 Cergy-Pontoise cedex, France}
\affiliation{Istituto dei Sistemi Complessi, CNR - Consiglio Nazionale delle Ricerche, via Madonna del Piano 10,
I-50019 Sesto Fiorentino, Italy}

\begin{abstract}
The activity of a sparse network of leaky integrate-and-fire neurons is carefully revisited with reference to 
a regime of a {\it bona-fide} asynchronous dynamics. The study is preceded by a finite-size scaling analysis, carried out to
identify a setup where collective synchronization is negligible. 
The comparison between quenched and annealed networks reveals the emergence of substantial differences when
the coupling strength is increased, via a scenario somehow reminiscent of a phase transition.
For sufficiently strong synaptic coupling, quenched networks exhibit a highly bursting neural activity, well reproduced by a self-consistent
approach, based on the assumption that the input synaptic current is the superposition of independent renewal processes. 
The distribution of interspike intervals turns out to be relatively long-tailed; a crucial feature
required for the self-sustainment of the bursting activity in a regime where neurons operate on average (much) below threshold.
A semi-quantitative analogy with Ornstein-Uhlenbeck processes helps validating this interpretation. 
Finally, an alternative explanation in terms of Poisson processes is offered under the additional
assumption of mutual correlations among excitatory and inhibitory spikes.
\end{abstract}

\maketitle

\section{Introduction}

The characterization of the spiking activity of neuronal networks is a long standing problem even with reference to
the asynchronous regime: simple from a dynamical point of view, but extremely relevant for understanding 
cortex dynamics~\cite{ecker2010,bal2}. A moment's reflection indeed suggests that this is not a trivial task 
whenever the self-generated neuron-input current is not constant: to what extent can the fluctuations be 
treated as a stochastic process?

One of the most popular models used to study neural dynamics consists of two coupled populations of 
excitatory and inhibitory leaky integrate-and-fire (LIF) neurons accompanied by refractoriness and 
delay~\cite{brunel2000,ostojic}. Incidentally, this model was also proposed for the characterization 
of asynchronous dynamics in balanced networks~\cite{bal1}. 
In a ground breaking paper, Brunel~\cite{brunel2000} proposed to treat the input current as a 
$\delta$-correlated Gaussian process~\cite{ricciardi}, thereby deriving and solving analytically 
a self-consistent Fokker-Planck equation.
Although this approach turns out to be quantitatively accurate for relatively small coupling strengths, 
the same is no longer true for stronger coupling (see~\cite{Ullner_Chaos2018}), when large deviations 
from the theoretical predictions are observed.
These deviations may in principle originate from various sources: 
(i) the spontaneous onset of irregular collective dynamics, 
which has been found even for relatively small network connectivities~\cite{Ullner_Chaos2018,Politi_epjst_2018};
(ii) the non Poissonian nature of the spiking activity; 
(iii) large amplitude of the single spikes and the consequential possible failure of a perturbative, linear
approach; (iv) the presence of non-negligible finite-time correlations.

Several alternative approaches have been indeed proposed. For instance an exact treatment of 
shot-noise, for spike amplitudes not vanishingly small, which leads to a 
mixed Fokker-Planck/master-equation formalism (see~\cite{Richardson_prl_2010,olmi2017}). Unfortunately, we are not aware 
of any way to make the approach self-consistent, by inferring the input properties on the basis of the observed output. 
Anyhow, since this approach assumes a Poissonian distribution of the inter-spike-intervals -- a property
largely unsatisfied for large synaptic coupling -- one should anyway look for different approximation schemes.

A different strategy was proposed by Dummer {\it et al.}~\cite{dummer2014}, based on the self-consistent 
derivation of the power spectrum of the spiking activity. 
The advantage of this method is that no assumption is made on the spectral shape of the synaptic current.
While the original implementation proved unstable already for relatively small coupling strengths, the variant
recently proposed in~\cite{pena2018} leads to seemingly accurate reproduction of the network dynamics.
We shall treat it as a reference for some of our considerations.

In this paper we revisit the problem, starting from the accuracy of numerical simulations and the presence of 
finite-size corrections.
In Refs.~\cite{Ullner_Chaos2018,Politi_epjst_2018}, it was indeed shown that the firing activity of a network
of 10,000 neurons with a in-degree $K=1,000$ is strongly affected by the presence of collective
irregular dynamics. Our first goal has been therefore that of finding the minimal network-size such that
collective effects are negligible.

The first result is that quenched networks (characterized by a fixed adjacency matrix) exhibit a
substantially different behaviour from annealed ones (where post-synaptic neighbours are randomly selected 
whenever a spike is emitted). Quite interestingly, the difference emerges almost abruptly above a ``critical"
synaptic coupling strength (namely, $J \approx 0.25$). 
The existence of two seemingly different phases was already claimed by Ostojic~\cite{ostojic}, but challenged in
Ref.~\cite{comment}. Here, we do not investigate the behaviuor in the vicinity of the hypothetical phase transition,
but rather focus on the characterization of the large-coupling regime, since we believe that the accurate
characterization of a given phase has higher priority.

One of the main results of this paper is that the synaptic current can be accurately represented as
the superposition of independent identical renewal processes, each characterized by a suitable  interspike-interval 
(ISI) distribution. We also show that the correlations due to the long-tailed ISI distribution can be equivalently 
represented as long-term memory in the symbolic representation of inhibitory vs excitatory spikes. 

More precisely, in Sec~\ref{sec:mm}, we introduce the model and define the indicators later used to characterize 
and discuss the various dynamical properties. In the following Sec~\ref{sec:nd} we illustrate the firing activity 
of the quenched network, computing several indicators for different coupling strengths. 
A relatively quick discussion is also devoted to the annealed set-up to show the differences
with respect to the quenched case.  
In Sec.~\ref{sec:sc} we first introduce the two self-consistent approaches herein implemented to characterize 
the neural activity. The former one, based on the distribution of ISIs, provides 
a rather accurate description. The latter, already proposed in~\cite{dummer2014,pena2018}, reveals an
unexpectedly stable fixed point, which, however, is further away from the results of accurate simulations.
In Sec.~\ref{sec:chara}, we turn our attention to the bursting activity observed 
for large coupling in the attempt of explaining how neurons operating on average below threshold
are able to exhibit a strong firing activity. 
Finally, in Sec.~\ref{sec:cop} we summarize the main results and focus on the still open problems.

\section{Model and Methods}
\label{sec:mm}

\subsection{Network Model}

Due to its relevance in the context of asynchronous dynamics in balanced networks
~\cite{brunel2000,ostojic, ullner2016, politi2017} we consider the following sparse spiking network  
of LIF neurons. The network is composed of $bN$ excitatory and $(1-b)N$ inhibitory cells;
the membrane potential $V_i$ of the $i$-th neuron evolves according to the equation
\begin{equation}
\tau \dot V_i = R (I_{0}+I_i) - V_i  \; ,
\label{eq:LIF}
\end{equation}
where $\tau = 20$ ms is the membrane time constant, $R I_{0} = 24$~mV is an external DC supra-threshold ``current", while $R I_i$ is
the synaptic current arising from the recurrent coupling
\begin{equation}
RI_i(t) = \tau J \sum_n G_{ij(n)} \delta(t-t^{(j)}_n-\tau_d) \; ,
\label{eq:general}
\end{equation}
where $J$ is the coupling strength and the sum runs over all the spikes
emitted at time $t^{(j)}_n < t$ from the pre-synaptic neurons $j(n)$ connected to neuron $i$.
 $G_{ij}$ is the adjacency matrix and
its elements assume the following values: $G_{ij}=1$ ($-g$), if the pre-synaptic neuron 
$j$ is excitatory (inhibitory), otherwise $G_{ij}=0$.
If $V_j$ reaches the threshold $V_{th} = 20$~mV at time $t^{(j)}_n$, two events
are triggered: (i) the membrane potential $V_j$ is reset to $V_r = 10$~mV and held fixed 
for a refractory period $\tau_r=0.5$ ms; (ii) a spike is emitted and received $\tau_d = 0.55$ ms 
later by the post-synaptic cells connected to neuron $j$ according to $G_{ij}$. 
Except for the system size $N$, all parameters are set as in Ref.~\cite{ostojic}: 
$b=0.8$, $K=1000$, and $g=5$, so that each neuron receives input from $b K$ ($(1-b)K$) excitatory (inhibitory) pre-synaptic neurons.  
Besides this {\it quenched} setup, we have considered {\it annealed} networks, where the 
post-synaptic neighbours are randomly chosen at each spike emission. 
As a matter of fact, in the former (latter) case, the in-degree
(out-degree) is equal to $K$ for each neuron, while the out-degree (in-degree) is binomially 
distributed with average $K$ and standard deviation $\sqrt{K}$, in the thermodynamic limit.  
Our choice was dictated by efficiency of the numerical simulations; however we have verified
that no substantial changes are observed if, instead of fixing the number of links equal to $K$,
the probability of each link is set equal $c=K/N$, as in truly Erd\"os-Renyi networks.

\subsection{Methods}

A detailed description of network dynamics requires looking both at the microscopic and the macroscopic level.

\subsubsection{Microscopic Indicators}

The dynamics of a spiking neuron is usually characterized in terms
of the probability distribution function (PDF) $Q(T)$ of the ISIs $T$ and of the associated moments:
namely, the mean ISI $\overline T$ and the standard deviation $\sigma_T$ of $T$.
Usually, the regularity/irregularity of the dynamics is 
quantified by the so-called coefficient of variation,
\[
C_v =   \frac{\sigma_T}{\overline T}  \; ,
\]
equal to zero for a periodic dynamics and to 1, for Poissonian spike trains.
It should be noticed that $C_v$ can be larger than 1 for the so-called bursting
dynamics, when the neuron alternates periods of silence and high activity.
The firing rate of a neuron is simply given by $\nu = 1/{\overline T}$.
In order to characterize the network activity we estimate the mean coefficient of variation 
$\langle C_v \rangle$ 
and the mean firing rate 
$\langle {\nu} \rangle$, where $\langle \cdot \rangle$ represents an ensemble average over all neurons.

An important observable is the power spectrum $S(f) = \langle |\tilde u(f)|^2 \rangle$, where 
$\tilde u(f)$ with $f=m/(M\delta t)$ is the Fourier transform of the neural activity $u(t)$, determined 
by computing the number of spikes emitted in $M$ consecutive time intervals of duration $\delta t$.
Further observables we focused on are the phase correlations among different frequencies $f$ and $h$, which 
can be quantified by the normalized indicator,
 \begin{equation}
 D(f) = \sqrt{\frac{\sum_{h \neq f} \langle | \tilde{u}(h) \, \tilde{u}^*(f)|^2\rangle}{W \, S(f) }
}
\label{phase_corr}
\end{equation}
where $W = \langle \sum_h |\tilde{u}(h)|^2 \rangle$ is the total power of the spectrum.
One can check that $0\le D(f)\le 1$, the lower (upper) bound corresponding to uncorrelated 
(perfectly correlated) channels. 
The typical values we have used in our simulations are $\delta t = 0.11$ ms and
$M=2^{15}$ ($M=2^{12}$) for the power spectrum (phase correlations) estimation.

\subsubsection{Macroscopic Indicators}

At the mean-field level, the network evolution is captured by
the instantaneous PDF $P(v,t)$ of the membrane potentials of the neurons. In the limit case of
an infinitely large in-degree, the perfectly asynchronous regime is characterized by a
constant firing rate $\langle {\nu} \rangle$ \cite{gerstner}. 
This implies that the flux of neurons along the $v$-axis is independent of both time
and potential value, i.e. the corresponding PDF  $P_0(v)$ should be stationary.  

Deviations from stationarity reveal the presence of a collective dynamics.
In order to measure the level of coherence in the network
dynamics, a commonly used order parameter is~\cite{scholarpedia}
\begin{equation}
\rho^2 \equiv \frac{\overline{\langle v\rangle^2}-\overline{\langle v\rangle}^2}
    {\langle \overline{v^2}-\overline{v}^2\rangle} \; ;
    \label{rho}
\end{equation} 
where the overbar denotes a time average.
In practice, $\rho$ is the rescaled amplitude of the standard deviation of the average $\langle v\rangle$.
When all neurons behave in exactly the same way (perfect synchronization),
the numerator and the denominator are equal to one another and $\rho=1$. If instead, they are
independent as in an asynchronous regime, $\rho \approx 1/\sqrt{N}$ due to the central limit theorem. \\

\section{Network Dynamics}
\label{sec:nd}

\subsection{The quenched network}
\begin{figure}
\begin{center}
\includegraphics*[width=0.48\textwidth]{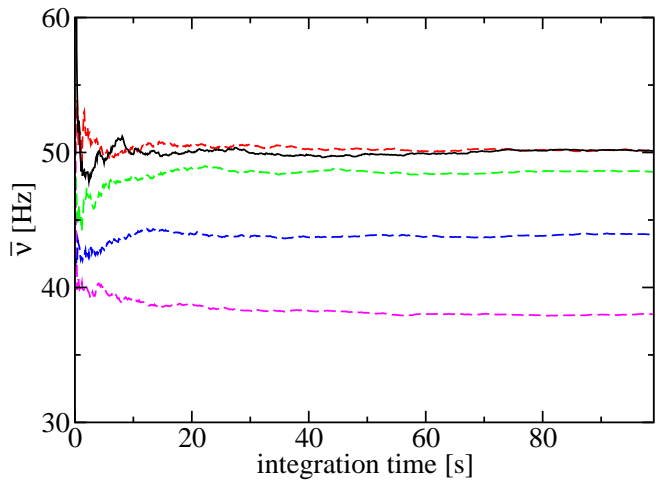}
\end{center}
\caption{Running average of the firing rate in a network of $N=10^5$ neurons with a coupling strength $J=0.8$ and overall connectivity
$K=10^3$. The black solid line refers to event-driven simulations; the dashed lines correspond to Euler integration scheme with different
time steps ($\delta t =$ 0.1 ms, 0.04 ms, 0.01 ms, 0.001 ms, from bottom to top).}
\label{fig:frate}
\end{figure}

Our simulations have been mostly performed by implementing an exact event-driven scheme;
see~\cite{Politi_epjst_2018} for a description of the details. Since, however, some simulations required
implementing a clock-driven Euler scheme and since this latter approach is often used in the literature, we
have first compared the two algorithms for a network of $10^5$ neurons with a coupling strength $J=0.8$.
From the results, reported in Fig.~\ref{fig:frate}, we see that the time step $\delta t$ used in the implementation
of Euler's algorithm should not be larger than $10^{-3}$ ms in order to get results essentially in agreement
with the event driven scheme. This is indeed the value employed in our simulations performed with the
Euler's scheme. Notice that for $\delta t= 0.1$ ms, a value often chosen in the literature, the firing rate is substantially underestimated
(by approximately 24\%).
 
\begin{figure}
\begin{center}
\includegraphics*[width=0.48\textwidth]{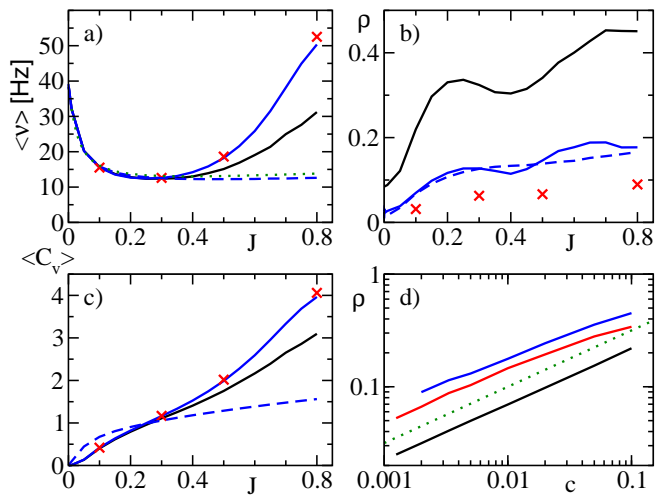}
\end{center}
\caption{Collective properties of the
network dynamics versus the coupling strength $J$: (a) average firing rate $\langle \nu \rangle$;
(b) the coherence order parameter $\rho$; (c) the mean coefficient of variation $\langle C_v \rangle$.
Solid lines and symbols refer to quenched networks (namely, black lines correspond to $N=10^4$; blue ones to $N=10^5$ and (red) crosses to $N=8\times 10^5$) (blue) dashed lines correspond to simulations of the annealed
network performed for $N=10^5$ neurons. The (green) dotted line in panel (a) corresponds to the
theoretical prediction $\nu_T$ derived in \cite{ricciardi,brunel2000}. Finally, panel (d) displays the coherence order parameter 
$\rho$ versus the connectivity $c=K/N$, reporting only results for the quenched network for 
different synaptic coupling: namely, $J=0.1$, $J=0.5$, and $J=0.8$ (solid lines from the bottom to the top). 
The dotted line refers to the theoretical scaling law $\sqrt{c}$.}
\label{network_Jmeasures}
\end{figure}

We then proceed to analysing the dependence of the average firing rate ${\langle \nu \rangle}$ on the coupling strength
$J$. The black, solid curve in Fig.~\ref{network_Jmeasures}(a) has been obtained for $N=10^4$
and exactly the same parameter values as in Ref.~\cite{ostojic}. As reported therein,
$\langle \nu \rangle$, after an initial drop, increases with the coupling strength $J$.  
One of the goals of this paper is to understand the origin
of this growth in a network where inhibition is expected to prevail over excitation.
A theoretical estimate $\nu_T$ of the average firing rate in the asynchronous regime of a sparse network can be derived from  the stationary solution of a self-consistent Fokker-Planck equation, under the assumption of an uncorrelated Poissonian activity of the neurons~\cite{ricciardi,brunel2000}. However, this prediction, based on the diffusion approximation~\cite{ricciardi} and reported as a dotted green curve in Fig.~\ref{network_Jmeasures}(a), 
is able to reproduce only the initial part of the curve $\langle \nu(J) \rangle$, 
while it fails to describe the growth observable for larger coupling. 
Furthermore, in Refs.~\cite{Ullner_Chaos2018,Politi_epjst_2018} it was found that the corresponding dynamical 
phase is far from asynchronous; this is testified by the behavior of $\rho(J)$, reported in 
Fig.~\ref{network_Jmeasures}(b), where we can see that the order parameter $\rho$ can be as large as 0.5.
Considering that the theoretical prediction has been derived under the assumption of a
strictly asynchronous regime (i.e. $\rho =0$), it is therefore crucial to separate out 
the effects of the collective dynamics.

This can be done by increasing the network size, while leaving the  in-degree fixed
(namely, $K=1000$). Quite surprisingly, the firing rate obtained for $N=10^5$,
(see the upper blue solid curve in Fig.~\ref{network_Jmeasures}(a)) 
displays an even more pronounced growth than for $N=10^4$, in spite of a weaker synchronization, as
shown in Fig.~\ref{network_Jmeasures}(b). 
The analysis reported in Fig.~\ref{network_Jmeasures}(d), where $\rho$ vs. $c=K/N$ is 
reported for three different synaptic coupling values, shows that
the collective effects increase as $\rho \approx \sqrt{c}$, consistently with the theoretical 
expectations~\cite{brunel2000}. 

Going back to Fig.~\ref{network_Jmeasures}(a), we see that upon further increasing $N$ above $10^5$, 
the firing rate for a given coupling strength $J$ saturates to a finite value.
Altogether, we can safely conclude that the increase of $\nu$ for $J \gtrsim 0.3$
is a genuine property of a bona fide asynchronous activity and it should be explained as such.
Our simulations suggest that the system-size $N=10^5$ is large enough to ensure nearly asymptotic 
results and small enough to allow for affordable simulation times.
From now on, all simulations will refer to this network size, unless stated otherwise.
The main questions we want to address are understanding: (i) the features of such a high firing-rate regime 
and (ii) why it deviates so strongly from the diffusion approximation \cite{ricciardi,brunel2000} 
even for a not-too-large synaptic coupling $J$ in a setup where correlations 
among the different neurons are practically absent. 

Before proceeding along these lines, it is useful to provide a more detailed description of the network activity.
In Fig.~\ref{network_Jmeasures}(c) the mean coefficient of variation $\langle C_v \rangle$ is plotted 
versus $J$ for different networks sizes.
There, we see that $C_v$ steadily increases with $J$ and converges to some asymptotic value upon decreasing $c$.
In practice, the neural activity can be never treated as a Poissonian process, as requested
by the diffusion approximation employed in \cite{brunel2000}; it is either more regular (for small coupling),
or substantially more intermittent (as for $J>0.4$).
Therefore, it should not come as a surprise that a theoretical approach, such as that in 
Ref.~\cite{ricciardi,brunel2000}, based on the assumption that $\langle C_v \rangle=1$, is not accurate. 

\begin{figure}
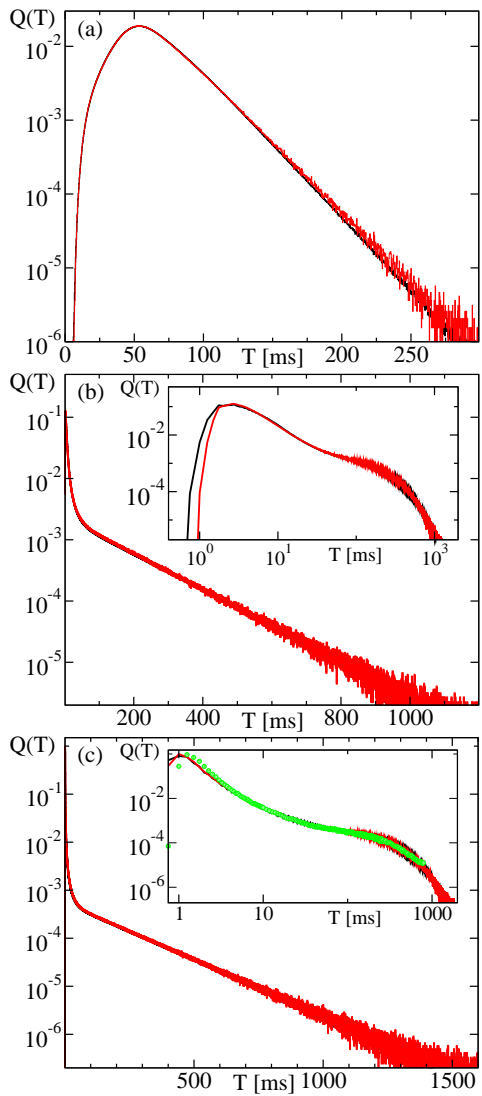

\begin{center}
\includegraphics*[width=0.35\textwidth,clip]{Fig3a.eps}
\includegraphics*[width=0.35\textwidth,clip]{Fig3b.eps}
\includegraphics*[width=0.35\textwidth,clip]{Fig3c.eps}
\end{center}
\caption{The PDF $Q(T)$ of the ISIs $T$ generated by the full quenched network (black) 
and as obtained after the first step of the renewal process recursive procedure (red curve).
Panels (a), (b) and (c) refer to $J=0.1$, $J=0.5$, and $J=0.8$, respectively. The insets contain the
same information in log-log scales to emphasize the initial quasi-power-law decay. The inset in panel (c)
shows (in green) the output of a single neuron by a symbol-correlated Poisson process (see Sec.~\ref{sec:chara} C).}
\label{ISI}
\end{figure}

$C_v$ gives only a rough information about the distribution of the ISIs. 
It is worth turning our attention to the full shape of the ISI distribution $Q(T)$.
In Fig.~\ref{ISI}, we plot $Q(T)$ for $J=0.1$, $J=0.5$ and $J=0.8$;
in all cases, we see that for large enough ISIs, the PDF is characterized by an exponential tail 
as for a Poissonian process. However, for small ISIs, the PDF is substantially different.
For weak coupling, very small ISIs are strongly inhibited~\footnote{One should also remember that because of
the refractory period, $T>\tau_R$}:
this is an obvious consequence of the nearly  constant input current (mean driven).
For stronger couplings, the PDF exhibits a quasi power-law decay which extends up to 10-20 ms (see the insets
of Fig.~\ref{ISI}(b) and (c)). These features are consistent
with the characterization of the ISI distributions reported in \cite{ostojic2011} for spiking neurons 
driven by fluctuating inputs. In particular, the PDF shown in Fig.~\ref{ISI}(a) is expected to 
emerge when the average effective input current (including the contribution of the synaptic coupling)
lies between the threshold and the reset value, which is indeed the case.

\begin{figure}
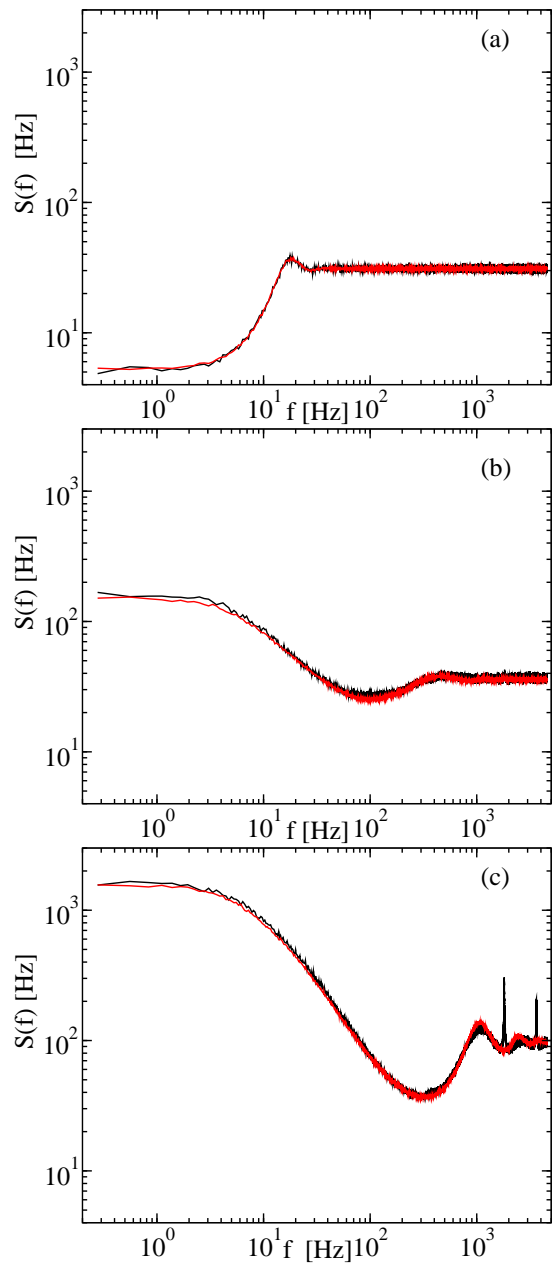

\begin{center}
\includegraphics*[width=0.4\textwidth]{Fig4a.eps}
\includegraphics*[width=0.4\textwidth]{Fig4b.eps}
\includegraphics*[width=0.4\textwidth]{Fig4c.eps}
\end{center}
\caption{Power spectra $S(f)$ of the neural activity. Black curves refer to the full quenched 
network for $N=10^5$; red curves refer to the first step of the renewal process approximation. 
Panels (a), (b), and (c) refers to $J=0.1$, $0.5$, and $0.8$, respectively.}
\label{power}
\end{figure}

Furthermore, we have computed the power spectrum $S(f)$ of single spike sequences. 
For weak coupling ($J=0.1$), a small peak is visible at $f = 17~Hz$ in Fig.~\ref{power}(a)~\footnote{Here and
everywhere power spectra are represented, they are normalized in such a way that the total power is 
obtained by integrating over all positive frequencies: $0\le f< +\infty$.}
it is reminiscent of the periodic activity of the uncoupled neurons. 
At higher frequencies, the spectrum is practically flat, i.e. it is approximately white.
Upon increasing the coupling, the spectrum starts exhibiting a low-frequency peak
suggesting the presence of ``long"-time correlations. This feature will be further discussed in Sec.~\ref{sec:chara}
with reference to the emergence of a bursting activity.
For $J=0.8$ subsidiary peaks, related to the delay, emerge for $f=1818$~Hz and its multiples. 
The delay is always present but for unexplained reasons 
pops up only for large coupling when the white-noise background is even larger.
 
\subsection{The annealed network}

So far we focused on the dynamical properties of a network
characterized by a quenched distribution of synaptic connections.
However, all theoretical approaches developed to characterize the firing activity do not take into account
the actual, invariant structure of the connections. Even more, theoretical approaches do not include delay at all.
Therefore, it is natural to ask to what extent the quenched nature of the network may be considered responsible
for the observed asynchronous activity. This question can be addressed by considering an
annealed network, where the ``neighbours" of each given neuron are randomly assigned each time a spike is emitted.
More precisely, we proceed as follows: whenever a neuron fires, we still assume that the quality of the 
spike (excitatory vs inhibitory) is determined by the neuron itself, but we randomly choose $K$ receiving neurons 
regardless of their quality. Moreover, we exclude self-connections, i.e. the sender must differ from the receiver. 
Finally, we keep all parameters as in the quenched network. This guarantees that on average each neuron receives
$b K$ excitatory inputs and $(1-b) K$ inhibitory ones.

The numerical results for the average firing rate are reported in Fig.~\ref{network_Jmeasures}(a). There 
we observe a good agreement with the behavior of the quenched network for $J \lesssim 0.3$, while 
increasing deviations emerge for larger coupling strengths. 
Interestingly, the behavior of the annealed setup is very close to the theoretical prediction $\nu_T$ \cite{brunel2000}. 
This is not entirely surprising, since, as anticipated, the theoretical approach implicitly assumes an annealed
connectivity. An additional justification for this agreement is the $C_v$-values reported in 
Fig.~\ref{network_Jmeasures}(d) (see the dashed curve) which are much smaller than in the quenched case 
and closer to 1, the value corresponding to a Poisson process.

\section{Self-consistency}
\label{sec:sc}

In the previous section we have seen that quenched and annealed networks behave in a substantially different way,
when the coupling strength is larger than $J=0.25$.
To what extent is this difference the signature of the crucial role played by a fixed structure of 
the synaptic connections? 

In this section we address this issue by implementing a self-consistent approach, where the input current
is assumed to be the superposition of independent signals, each sharing the same ``statistical'' properties
of the single-neuron activity.
Two different approximation schemes are hereby discussed: (i) the hypothesis of a perfectly renewal process (RP);
(ii) mutually uncorrelated frequency channels (also termed Gaussian approximation).
Here below we show that the former one provides a more accurate representation of the neural activity.

\subsection{Renewal process}

A renewal process is fully characterized by the ISI probability distribution $Q(T)$.
Assuming $Q(T)$ is known, a typical spike sequence can be readily generated by randomly drawing 
a series of $T$ values accordingly to this distribution.
At variance with Ref.~\cite{dummer2014}, where the authors suggested the idea of approximating the 
synaptic current with a renewal process, here we limit ourselves to assume that the single-neuron output activity 
is an RP, but we do not extend the assumption to the input, which is treated as the superposition of $K$ independent RPs. 
This is an important difference since, as already remarked in~\cite{Lindner2006}, the superposition of RPs 
is not renewal itself unless the single processes are purely Poissonian (this is not our case). 
So, at variance with Ref.~\cite{dummer2014}, we relax the condition of a strictly renewal input process and 
replace this Ansatz with the more general hypothesis of a superposition of independent RPs.

In practice, we have implemented the following recursive procedure: 
given the ISI distribution $Q_k(T)$ determined in the $k$th step,
we have generated the synaptic current $R I$ of a generic neuron 
(in the $(k+1)$st step) by superposing $K$ independent RPs all built according to the same distribution
$Q_k(T)$ (under the constraint that $bK$ spikes are excitatory and the remaining ones inhibitory). 
Upon afterwards
integrating the single-neuron equation, we have generated the firing activity induced by the current $R I$,
thereby determining the $(k+1)$st distribution $Q_{k+1}(T)$. 

\begin{figure}
\begin{center}
\includegraphics*[width=0.45\textwidth]{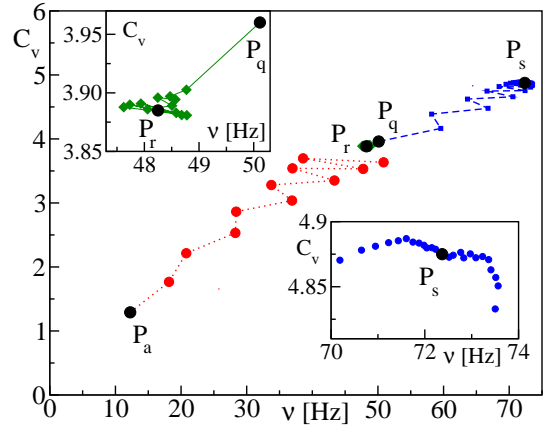}
\end{center}
\caption{Results of the recursive procedure based on the renewal process and the Gaussian approximation. 
$P_q$ and $P_a$ identify the dynamics of quenched and annealed networks, respectively. 
Red circles (green diamonds) refer to the iterates of the RP iterative procedure while starting from $P_a$
($P_q$); green diamonds are better visible in the enlargement reported in the upper inset, where $P_r$ 
denotes the fixed point of the renewal process approximation. Blue squares refer to the iterative process based 
on the Gaussian approximation:
$P_q$ is the initial condition, while $P_s$ denotes the fixed point of this approach (see the lower inset
for an enlargement of the later stages).  All data have been obtained for $J=0.8$).}
\label{iterate}
\end{figure}

We first focused on $J=0.8$, since the theoretical prediction $\nu_T$~\cite{ricciardi} is significantly inaccurate 
for this coupling strength.
The initial condition $Q_0(T)$ has been selected as the distribution generated by a quenched network 
of $N=10^5$ neurons with an in-degree $K=1,000$. 
The corresponding firing rate is $\langle \nu \rangle = 50.4$~Hz and its coefficient of variation is
$\langle C_v \rangle=3.97$ ~\footnote{In finite networks, sample-to-sample fluctuations are expected. 
Simulations of five different networks show that the standard deviation of 
$\left < \nu\right >$ is $\sigma_\nu \approx 0.4$, while that of $C_v$ is $\sigma_C = 0.01$.
Additionally, one expects the single steps of the recursive procedure do be affected by statistical
fluctuations: we have verified that the uncertainty of $\left < \nu\right >$ is about 0.05, while that of
$C_v$ is approximately $0.005$.}.
This pair of values is represented by the point $P_q$ in Fig.~\ref{iterate}: it corresponds to the 
projection of the asynchronous state of the quenched network in this two-dimensional space. The iterates of the recursive procedure have been projected on the same plane; they are so close to each other
to be hardly discernible in the main panel (see the enlarged plot presented in the upper inset of
Fig.~\ref{iterate} for a clearer representation).
The closeness among consecutive iterates is confirmed by the shape of the ISI distribution: 
in Fig.~\ref{ISI}(c), we see that $Q_1(T)$ is practically indistinguishable from $Q_0(T)$.
Altogether, these observations strongly hint at the existence of a fixed point of the RP recursive 
procedure in the vicinity of $P_q$.
Further iterates start separating from each other, suggesting that the fixed point is a saddle, 
which initially attracts the trajectory along the stable manifold and eventually drives it away along the 
unstable manifold. 
If, in analogy to what done in Ref.~\cite{pena2018} for the Gaussian approximation, we include memory effects 
by building the new ISI PDF as the average of the last two distributions, the saddle is stabilized: the
resulting fixed point is represented in Fig.~\ref{iterate} as $P_r$.
The non perfect correspondence between $P_q$ and $P_r$ may have a double rationale: 
the RP assumption is not exact; the network size used to determine $P_q$ is not large enough. 

In order to test the quality of the RP approximation, we have studied the correlations of the 
sequence $T_n$ of consecutive ISIs, by estimating the so-called serial correlation coefficient~\cite{schwalger2008}
\begin{equation}
   C(m) = \frac{\langle T_{n+m}T_n \rangle - \langle T_n \rangle^2}{\langle T_n^2\rangle-
\langle T_n\rangle^2} \; .
\end{equation}
In a strictly renewal process $C(m)=0$ for $m\ge 1$. Tests made on the neurons of a quenched network for 
$N=10^5$ show that $C(1)$ is at most of order $10^{-3}$, suggesting that the neural activity is well approximated by an RP. 
On the other hand, since the order parameter $\rho$ is still relatively large for $N= 10^5$ 
($\rho = 0.17$), finite-size affects are probably the predominant source of differences between $P_q$ and $P_r$.

What if the same recursive procedure is applied, starting from the dynamical regime exhibited by the annealed 
network (see the point $P_a$ in Fig.~\ref{iterate}).
Forward iterates rapidly moves away from $P_a$ and approach $P_r$ (see the full circles in Fig.~\ref{iterate}). 
The increasing amplitude of the ``transversal" fluctuations confirm that $P_r$ is a saddle point.
Furthermore, the recursive procedure shows that $P_a$ -- a fixed point of the annealed process -- 
is not a fixed point of the RP iterative procedure.
The reason is that while the temporal correlations exhibited by the single-neuron activity 
(encoded in the bursting activity) are preserved by the RP approximation, they are lost in the annealed setup
because of the random reshuffling of the synaptic connections. The separation between $P_a$ and $P_r$
implicitly suggests the important role played by the bursting activity as it will be confirmed in the
following section.

Finally, we have implemented the RP approach also for smaller $J$-values, always finding evidence 
of a weakly unstable fixed point (actually, the degree of instability decreases upon decreasing $J$). 
The resulting message is that the stable asynchronous dynamics exhibited by the quenched network is well reproduced
by an unstable fixed point of a recursive transformation based on the RP approximation.

\subsection{Power spectrum}

Fourier analysis offers the opportunity for additional verification of the validity of the RP approximation.
In Fig.~\ref{power}, we compare the power spectrum of the single-neuron activity after the first 
iterate obtained under the RP approximation~\footnote{No appreciable differences can be noticed while referring to the following iterates.}
(see the purple curve) with the spectrum exhibited by the quenched network. The agreement is quite good
for all of the three tested coupling strengths, the major discrepancy being the
absence of peaks at multiples of $\nu_d=1818$~Hz for $J=0.8$, 
which cannot be reproduced by the RP approximation, since the delay is not included in
such formulation.

\begin{figure}
\begin{center}
\includegraphics*[width=0.45\textwidth]{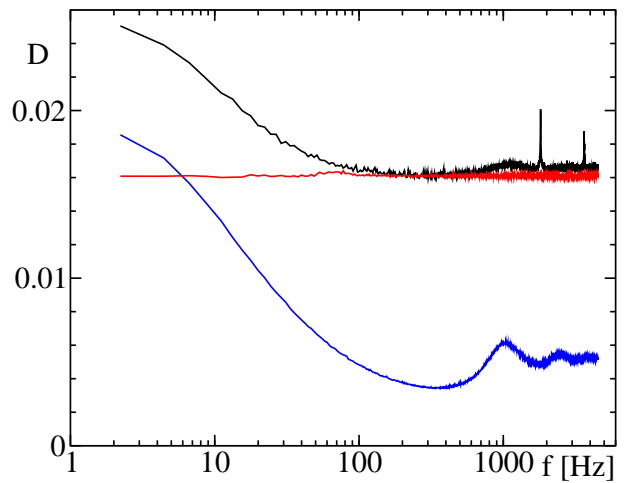}
\end{center}
\caption{Degree of phase correlations $D(f)$ of the Fourier transform of the spiking activity for $J=0.8$. 
The lowermost blue line corresponds to the RP; the uppermost black solid line corresponds to the activity of
the quenched network. Finally, the red line corresponds to the annealed network.}
\label{phasecor}
\end{figure}

Next, we have estimated directly the correlations among the phases of the Fourier modes, 
by computing $D(f)$ (see Eq.~(\ref{phase_corr})) both for the quenched network and the RP approximation 
(see upper and lower curves in Fig.~\ref{phasecor}, respectively). 
Phase correlations appear to be small in both cases (look at the vertical scale):
we attribute the larger amplitude exhibited by the quenched network to the presence of a 
residual collective dynamics, absent by definition in the RP approximation.
A comparably small level of correlations is found also in the annealed network (see the almost flat red line).

\begin{figure}
\begin{center}
\includegraphics*[width=0.45\textwidth]{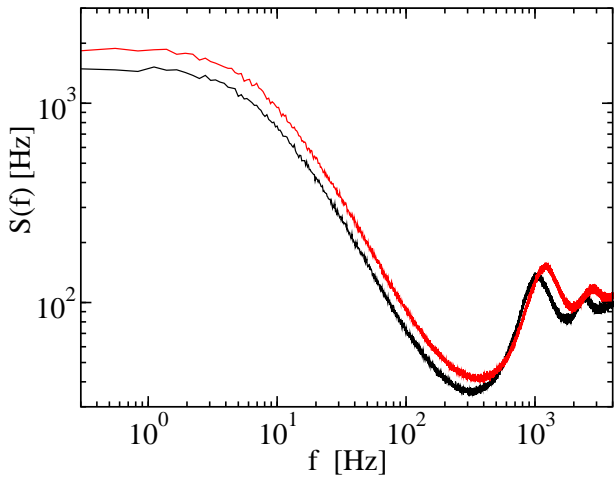}
\end{center}
\caption{Power spectra for the network activity.
The black curve corresponds to the spectrum of the activity of a single LIF neuron driven within the 
RP approach (it is basically indistinguishable from the true quenched network activity). 
The red curve is obtained by randomizing the phases of the input signal. Data refers to $J=0.8$.}
\label{phaserandom}
\end{figure}

As a second test of the relevance of phase-correlations, we have investigated the consequence of phase 
randomization within the RP procedure. More precisely, given the synaptic current $u(t)$ and its
Fourier transform $\tilde u(f)$, we have generated a new transform
$\tilde u_N(f) = |\tilde u(f)| \mathrm{e}^{i\phi(f)}$, by randomly assigning the phase $\phi(f)$ to the
frequency $f$. A new signal $u_N(t)$ is then obtained by back transforming $\tilde u_N(f)$.
The resulting spectrum of the firing activity of a neuron subject to the current $u_N(t)$ is 
presented in Fig.~\ref{phaserandom} (see the red curve). The difference with the original spectrum (see the
lower black curve) is not entirely negligible: it is around around $20\%$ in the low frequency region.

Finally, we have implemented the recursive procedure proposed in Ref.~\cite{dummer2014}, here briefly recalled.
Let $S_k^{(o)}(f)$ denote the power spectrum of the single-neuron spiking activity at the $k$-th recursive
step. Let also $S_{k+1}^{(i)}(f)$ denote the power spectrum of the synaptic current in the 
$(k+1)$st recursive step.
In the asynchronous regime, the synaptic current is the superposition of $K$ independent signals
($bK$ excitatory and $(1-b)K$ inhibitory), all characterized by the same spectrum $S_k^{(o)}(f)$. 
Taking into account the amplitude of the single spikes, we have that 
$S_{k+1}^{(i)} = [J^2K(b+(1-b)g)]S_k^{(o)}(f)$~\footnote{Leaving aside the zero-frequency channel which
contributes to the average and is treated differently.}.
The definition of the procedure is completed by adding the ``rule" to generate $S_{k+1}^{(o)}$, given
$S_{k+1}^{(i)}(f)$. This is done by feeding a single neuron with a phase-randomized spectrum (see the
paragraph above).
The self-consistent solution is finally identified by the condition $S_{k+1}^{(o)} = S_{k}^{(o)}$.

We have implemented this approach with an Euler integration step $\delta t = 10^{-3}$ms
starting from the initial condition $P_q$, the best proxy for the asynchronous regime.
The first 33 iterates are reported in Fig.~\ref{iterate},  where we see that they move away from
$P_r$ (see the blue squares) and approach a seemingly stable fixed point $P_s$. 
The relatively large difference between $P_s$ and $P_q$ suggests that this approximation scheme is not so accurate as the RP method and implicitly means that the phase correlations built while integrating the LIF equation are not negligible~\footnote{The quantitative differences with the results for  $P_s$ reported in Ref.~\cite{pena2018} are quite likely to be attributed to the lack of accuracy in the integration scheme 
employed therein.}.

\section{Bursting activity}
\label{sec:chara}
In the previous sections we have seen that for strong coupling the neural activity is characterized 
by a large $C_v$, a typical indicator of burstiness.
Here, we discuss more in detail the properties of this form of asynchronous regime, starting
from the basic question of how it is possible for it to be self-sustained.

\begin{figure}
\begin{center}
\includegraphics*[width=0.45\textwidth]{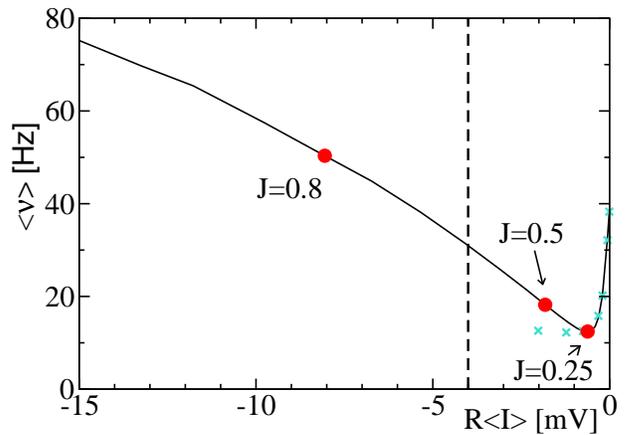}
\end{center}
\caption{Firing rate $\langle \nu \rangle$ versus the synaptic ``current" $R\langle I\rangle$  
for different coupling strengths for the quenched network (solid line and red circles). Zero coupling corresponds to the rightmost point, where the synaptic current
obviously vanishes. The vertical dashed line separates the region where neurons operate above (right) from
below (left) threshold. The green crosses report the outcome of the annealed network. The two crosses deviating from the solid line belong to $J=0.5$ and $0.8$ in the annealed setup.
}
\label{fig:nui0}
\end{figure}

In the asynchronous regime, the average input current induced by the synaptic coupling is
\[
\langle I \rangle = \frac{J}{R} (n_E-g n_I) = \frac{KJ}{R} \nu (b(g+1)- g))
\]
where $n_E$ ($n_I$) denotes the number of excitatory (inhibitory) spikes received per time unit.
Depending whether $R_0\langle I \rangle$ is larger or smaller than $V_{th}-RI_0=-4$, the neuron operates 
either above or below threshold. In fact, in the latter case, the velocity field crosses the zero axis below the 
threshold $V_{th}$, preventing threshold passing.
In Fig.~\ref{fig:nui0}, we plot the firing rate versus $\langle I\rangle$ for different values of the
coupling strength $J$ (see the solid line): increasing $J$ corresponds to moving leftward along the curve, starting from 
the rightmost point, which corresponds to the uncoupled limit. Upon increasing $J$, 
$\langle I \rangle$ decreases monotonically: this is the consequence of the prevalent inhibition. 
At the same time, the firing rate, after an initial drop, starts growing;
this happens for $J\approx 0.25$, as it can be inferred by comparing with Fig.~\ref{network_Jmeasures}(a).
The increase continues also when the neuron operates below threshold and surpasses the activity of the uncoupled
regime.

In the same figure, we report also the outcome of annealed-network dynamics (see the crosses): for
small $R\langle I\rangle$, i.e. for small coupling we see an almost perfect coincidence.
On the other hand, by further decreasing the internal current (i.e. upon increasing
the coupling strength), the firing rate does not increase in the annealed network, confirming
the qualitatively different behavior exhibited by the two setups. The annealed network operates above threshold.

\subsection{Correlations between membrane potential and synaptic current }

The counter-intuitive activity displayed by the quenched network requires an explanation. 
We have verified that the effective self-induced excitation is not the result of a symmetry breaking: 
all neurons (both excitatory and inhibitory) behave in the same way, as they should. 
More instructive information can be extracted by exploring the correlations between the actual value of the
membrane potential and the quality (excitatory vs. inhibitory) of the spike received by a given neuron. 
In other words, we have computed the relative fraction $s_E(V)dV$ of excitatory spikes received when
$V \in [V,V+dV]$.
If the receiving times were uncorrelated
with the membrane potential, then $s_E(V)$ would be independent of $V$ and equal to $b$.
Actually, this is expected within the framework of a $\delta$-correlated input signal as assumed
in~\cite{brunel2000}. 

\begin{figure}
\begin{center}
\includegraphics*[width=0.45\textwidth]{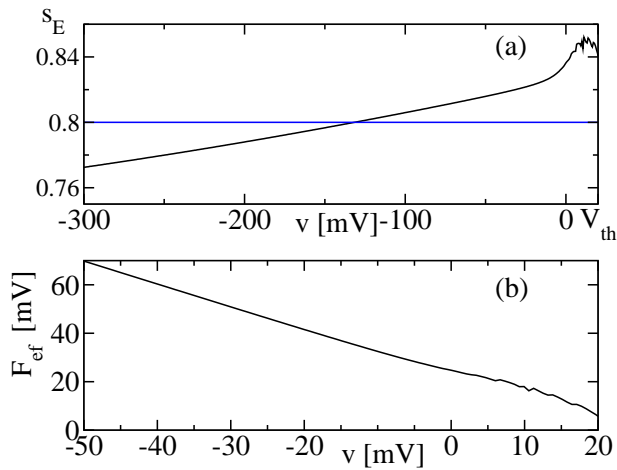}
\end{center}
\caption{(a) Fraction $s_E$ of excitatory spikes versus the current value $V$ for $J=0.8$. 
The horizontal line corresponds to the average global fraction $b=0.8$. 
(b) Effective velocity field $F_{ef}$ as from Eq.~(\ref{eq:feff}).}
\label{fig:ratioexcinh}
\end{figure}

Instead, in Fig.~\ref{fig:ratioexcinh}(a), we see sizeable deviations, 
especially in the vicinity of $V_{th}$, where $s_E$ is significantly larger
than $b$, hinting at a higher excitation than a priori foreseeable. We have verified that, as expected, 
the average of $s_E$ -- computed along the $V$-axis and weighted according to the stationary distribution $P(V)$)
 -- is equal to $b$ - see the horizontal line.

A perhaps more enlightening representation of the role played by the $V$-dependence of $s_E$ is proposed in 
Fig.~\ref{fig:ratioexcinh}(b), where we plot the effective velocity field
\begin{equation}
F_{ef} = RI_0 - V + K J \tau \langle \nu \rangle (s_E(1+g)-g) 
\label{eq:feff}
\end{equation}
where the $V$-dependent $s_E$ replaces $b$. Interestingly, the effective velocity field does not cross the zero
axis below threshold, showing that the neuron effectively operates above threshold, in spite of 
$R\langle I\rangle < V_{th}-RI_0$.
So, we can conclude that including the $V$-dependence of $s_E$ into the neural dynamics helps solving the
paradox of a neuron operating on average below threshold.
On a more quantitative level, we can interpret Eq.~(\ref{eq:feff}) as a deterministic evolution equation and
thereby compute the firing rate $\nu_a$ as the inverse of the time needed to reach the threshold $V_{th}$, while
starting from $V_r$ (augmented by the refractory time).
By inserting $\nu$- and $s_E(V)$-values obtained from the network simulations for $J=0.8$,
we find $\nu_a= 77$~Hz, to be compared with the observed rate $\nu = 50$~Hz. 
The agreement is not as good as one might had hoped for, but it should also be noted that
Eq.~(\ref{eq:feff}) does not account for the (strong) input fluctuations!

Moreover, the $V$-dependence of $s_E$ still needs to be understood.
Some light can be shed by arguing as follows. Let us introduce the joint probability
$P(E,L)$ that an excitatory spike reaches the neuron, when its membrane potential $V\in L=[V_0,V_{th}]$, where
$V_0$ is selected as the point where $s_E=b$,
The standard Bayesian inference rule implies that
\[
P(E|L) = \frac{P(L|E)}{P(L)} P(E)
\]
where $P(A|B)$ denotes the probability of observing $A$, given $B$; moreover $P(E)=b$, while
$P(L)$ is the probability of $V>V_0$, and
$P(E|L)$ is just the average of $s_E$ over $L$. 
Let us now focus on $P(L|E)/P(L)$: this is the probability of $V>V_0$ when an excitatory spike arrives
(rescaled to the unconditional probability to stay in $L$). 
If excitatory spikes arrive in bursts, for many of them the corresponding $V$-value is relatively 
large as a consequence of the excitation provided by the previous spikes. Therefore,  it is natural 
to expect $P(L|E)/P(L)>1$. This is precisely what we see in 
Fig.~\ref{fig:ratioexcinh}, where one can notice that $s_E$ is larger than $b$ close to threshold. Consistency then imposes that $s_E < b$ further away.

\subsection{Synaptic current: an Ornstein-Uhlenbeck process}

The role of correlations can be analysed from a different point of view: since the neuron is typically
under the action of a negative current, its membrane potential is kept away from threshold ($V<V_{th}$).
Only when relatively strong positive fluctuations of the input current arise, 
the neuron can overcome the threshold and emit a spike. 
If the correlations are sufficiently long-lasting, the fluctuation may stand long enough to allow for the 
emission of a sequence of spikes and give rise to a ``burst".
This mechanism has been already investigated in the past to quantify the spiking activity of a sub-threshold 
neuron subject to Ornstein-Uhlenbeck (OU) noise, finding that a long correlation time gives rise to a
bursting activity~\cite{moreno2004,schwalger2008}.
Unfortunately, we cannot make use of their formulas, since the correlation time is not much longer than
$\tau$. 
We can, nevertheless, proceed in a purely numerical way by approximating 
the input current $I$ with the outcome of an OU stochastic equation, namely
\[
\tau_c \dot I = \langle{I}\rangle -I + \xi  \; ,
\]
where $\langle I \rangle$ is the average current observed in the numerical simulations of the quenched
network, while $\tau_c$ is input correlation time and finally $\xi$ is a $\delta$-correlated white noise 
($\langle \xi(t')\xi(t'+t)\rangle = \sigma^2 \delta(t)$.
We have thereby tuned $\tau_c$ and $\sigma^2$, until the neuron activity is characterized by the expected
firing rate and the corresponding $C_v$. For
$J=0.8$, we have found $\tau_c \approx 160$ ms and $\sigma^2\approx 6.2 \cdot 10^{-2}$.
As a bonus, the resulting ISI distribution turns out to be quite similar to the expected one, 
the major difference being the peak which instead of being located in $T=1$ms (see the inset in Fig.~\ref{ISI}(c)),
occurs for $T\approx 4$ ms.
Altogether, one can nevertheless conclude that the OU approximation provides a reasonable description of the
input current.

We have implemented the same procedure for $J=0.5$: in spite of the similar bursting activity, the neuron operates on average above threshold and we have not found any way to parametrize the OU 
process so as to reproduce the observed activity.
On the other hand, a good reproduction of the neural activity is found for $J=1$, by assuming $\tau_c=145$ and
$\sigma^2=0.28$.
Two interesting comments are in order: (i) upon increasing the coupling, the correlation time does not increase: it
seems that $\tau_C \approx 140-160$ is an intrinsic property of the network; 
(ii) the noise amplitude increases by more than a factor 4, while passing from $J=0.8$ to $J=1$ and this is
the reason why the firing rate is larger in the latter case, even though the
neuron operates much more below threshold.
The increase of the effective noise can be attributed to two factors: a minor contribution comes from
the increased coupling strength (from 0.8 to 1); a more relevant contribution is the increased fluctuations of
the single-neuron activity quantified by the $C_v$. 

\subsection{Synaptic current: symbolic correlations}

We conclude this section by looking at correlations from a different point of view.
As shown in Sec.~\ref{sec:sc}, the output activity of the single neuron is well approximated by an RP, but we
do not expect the same to be true for the input, obtained by superposing $K$ independent such processes.  
In order to investigate the way correlations manifest themselves, we separately computed the ISI 
distribution of all excitatory and inhibitory spikes received by a given neuron.

\begin{figure}
\begin{center}
\includegraphics*[width=0.45\textwidth]{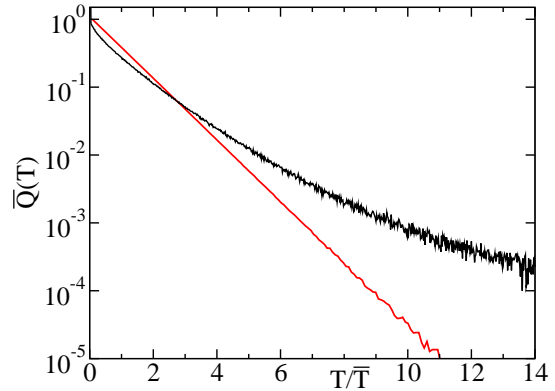}
\end{center}
\caption{The effective distribution $\bar{Q}(T)$ of inter-excitatory-spikes received by a single neuron for $J=0.8$. 
The black curve refers to the quenched network, the red one to the renewal process approximation.
The horizontal axis is rescaled by the mean firing rate of the respective population of excitatory spikes.}
\label{inISI}
\end{figure}

In Fig.~\ref{inISI} we report the ISI distribution of excitatory spikes 
(inhibitory spikes follow the same statistics) for both the original quenched network (black curve)
and within the RP approximation (red curve).
$\overline T$ represents the average separation between consecutive spikes, i.e. $\overline T$ is 
equal to the average single-neuron ISI divided by 800 -- the total number of incoming excitatory
synaptic connections. 
The red curve follows a clean Poissonian distribution, while the quenched network exhibits a slower than
exponential decay (in this time range); furthermore, in the latter case, the first channel is very large 
because of the unavoidable presence of avalanches occurring in the quenched setup (see~\cite{Politi_epjst_2018}).
We attribute most of the deviations from a pure exponential to the residual presence of collective dynamics.
In any case, this discrepancy is a minor issue: the relevant correlations are 
those between excitatory and inhibitory spikes, as revealed by the following test.
We have fed a single neuron with two different signals:
(i) a perfectly Poisson process composed of independent excitatory and inhibitory spikes; 
(ii) a synthetic signal built by assuming a Poisson distribution of consecutive spikes with the
same rate as the quenched network, 
but keeping the original symbolic ordering (i.e. $EEIEEEIE\ldots$, where the letter $E$/$I$ means that the
spike is either excitatory or inhibitory) observed in the quenched network.

\begin{figure}
\begin{center}
\includegraphics*[width=0.45\textwidth]{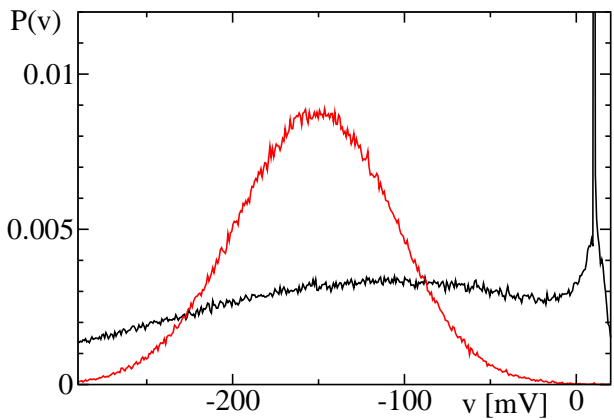}
\end{center}
\caption{Probability density of the membrane potential $P(v)$. The black curve exhibiting the divergence in $v=10$ 
corresponds to the synthetic signal described in text; the red curve is obtained by feeding the neuron with a purely
Poisson process with the same firing rate.}
\label{timeseries}
\end{figure}

The resulting membrane-potential distributions of the neuron are reported in Fig.~\ref{timeseries}.
The black curve, obtained by using the above mentioned synthetic signal, is very similar to the 
original distribution.  On a quantitative level, the resulting firing rate, the $C_v$ 
and the PDF of the ISIs are very close to the values exhibited by the RP approximation (deviations are smaller than 1\%):
see the inset of panel (c) in Fig.~\ref{ISI}. 
On the other hand, the red curve, originating from the strictly Poisson process is shifted towards 
very negative $v$-values and nearly vanishes close to the threshold, suggesting a very low 
firing activity as indeed observed.

Altogether this proves that the relevant correlations are contained in the symbolic 
representation of the spike sequence.

\section{Conclusions and open problems}
\label{sec:cop}

In this paper we have shown that upon increasing the coupling strength $J$ (and for $J>0.25$), a slightly inhibitory sparse
network of LIF neurons operates increasingly below threshold and yet fires at an increasingly fast rate.
This claim is supported by careful numerical simulations, tailored so as to marginalize the effects of collective
synchronization.

This counter-intuitive, self-sustained, activity observed in quenched networks, disappears in annealed networks,
i.e. in setups where the synaptic connections are continuously randomly reshuffled. In the latter case, the neural
activity is both weaker and more homogeneous (for $J=0.8$, the firing rate drops by a factor about 4).
The difference between quenched and annealed setup is reminiscent of replica symmetry breaking~\cite{spinglass},
but the anomaly of the phenomenon is mitigated by the observation that the quenched-network dynamics 
can be reproduced to a high degree of accuracy by an approach, the renewal-process (RP) approximation, 
which does not take into account the  structure of the synaptic connections.
Still, the comparison between quenched and annealed dynamics (see Fig.~\ref{fig:nui0} for the most 
enlightening representation)) seems to suggest the presence of a phase transition when $J$ is
increased. It looks like the two regimes deviate from one another above $J\approx 0.25$. 
This is reminiscent of the claim made by Ostojic about the existence of two distinct asynchronous
regimes \cite{ostojic}. This claim has been criticised in Ref.~\cite{comment}; we are also unable to find
evidence of a qualitative difference between the two regimes (above and below a supposedly critical point $J_c$).

Within the RP approximation, the neural activity is fully characterized by the ISI distribution.
In the limit of large coupling strengths, such a distribution exhibits a power-law decay, similar to what
found while studying the response of a single neuron to Ornstein-Uhlenbeck processes
\cite{moreno2004, schwalger2008} and similar to experimental observations made in the sensorimotor 
cortex of rats performing behavioural tasks \cite{tsubo2012}. It should, however, be noticed that 
in our case, the scaling range is much smaller than in the experimental observations.

\begin{figure}
\includegraphics*[width=0.45\textwidth,clip]{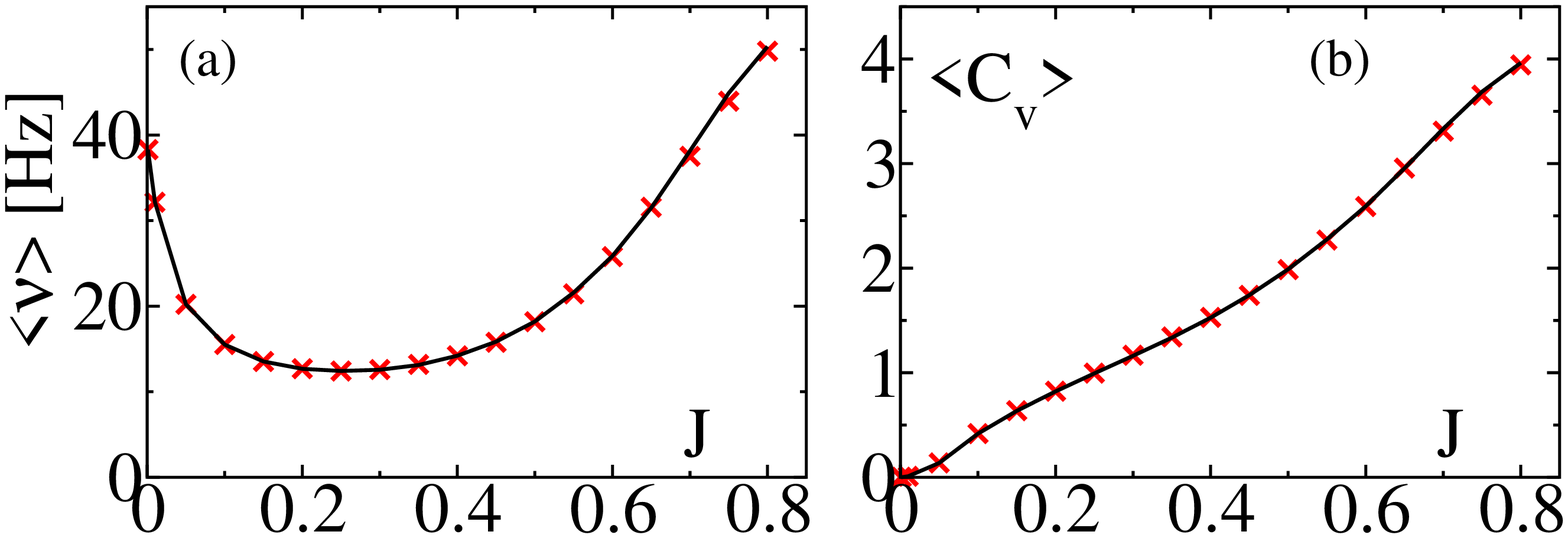}
\caption{Neural activity for different delay times. The solid curves all refer to 
$\tau_d=0.55$, while crosses refer to
$\tau_d=0.3$. All simulations are performed in a network with $N=10^5$ neurons and a connectivity $K=1000$.}
\label{fig:delay}
\end{figure}

The recursive process based on the RP approximation proves rather accurate in spite of not taking into account
the delay. It is therefore natural to ask whether this is true also in the quenched network. 
Simulations performed for different delay values confirm a substantial independence of the outcome on $\tau_d$.
In Fig.~\ref{fig:delay}, we compare the firing rate and the $C_v$ obtained 
for $\tau_d=0.3$ with the original simulations (performed for $\tau_d=0.55$). 

Finally, we wish to comment on the peculiar behavior of the network observed for large coupling strengths.
The strong firing activity is self-sustained by its burstiness (signalled by the large $C_v$ values), 
which, de facto, provides the relatively long correlations required to let a neuron below threshold fire. 
This clarifies the reason why the theoretical formula based on the assumption of $\delta$-correlated 
current fluctuations, fails to reproduce this regime. 
Recently, a more sophisticated self-consistent approach has been developed, where the Fokker-Planck equation
has been augmented to account for temporal correlations in the synaptic current~\cite{Vellmer_Lindner_prr2019}. 
Its (numerical) implementation to a weak-bursting regime looks promising.
It will be worth exploring its validity in a more inhibition dominated regime such as the one explored in this paper. 
Interestingly, the bursting activity is reproduced also assuming a strictly Poisson ISI distribution,
but retaining the correlations contained in the symbolic representation of the spike types 
(i.e. excitatory vs. inhibitory). A simple quantification of such correlations might open yet another route for
a quantitative characterization of the neural activity.

\acknowledgments
A.T. received financial support  by the Excellence Initiative I-Site Paris Seine (Grant No ANR-16-IDEX-008), by the Labex MME-DII (Grant No ANR-11-LBX-0023-01) (together with A.P. and E.U.) and by the ANR Project ERMUNDY (Grant No ANR-18-CE37-0014), all part of the French programme ``Investissements d'Avenir''. 
 


%

\end{document}